\documentclass[aps,prd,reprint,10pt,superscriptaddress,amsmath,showpacs,showkeys]{revtex4-1}
\usepackage{bm}
\usepackage{graphicx}
\usepackage{mathrsfs}
\usepackage{slashed}

\newcommand\de{\partial}
\newcommand\vek[1]{\bm{#1}}       
\newcommand\adj[1]{\overline{#1}} 
\newcommand\cc[1]{#1^{\mathcal C}}
\newcommand\he[1]{#1^{\dagger}}   
\newcommand\gr[1]{\mathrm{#1}}    
\newcommand\La{\mathscr L}
\newcommand\Ha{\mathscr H}
\DeclareMathOperator{\tr}{Tr}
\newcommand\eps{\epsilon}

\begin{document}

\title{Phase diagram of two-color quark matter at nonzero baryon and isospin
density}
\author{Jens O.~Andersen}
\email{jens.andersen@ntnu.no}
\affiliation{Department of Physics, Norwegian University of Science and
Technology, H{\o}gskoleringen 5,  N-7491 Trondheim, Norway}
\author{Tom\'{a}\v{s} Brauner}
\email{brauner@ujf.cas.cz}
\affiliation{Institut f\"{u}r Theoretische Physik, Goethe-Universit\"at,
Max-von-Laue-Stra{\ss}e 1, D-60438 Frankfurt am Main, Germany}
\altaffiliation{On leave from Department of Theoretical Physics,
Nuclear Physics Institute ASCR, CZ-25068 \v{R}e\v{z}, Czech Republic.}

\begin{abstract}
We investigate the properties of cold dense quark matter composed of two colors
and two flavors of light quarks. In particular, we perform the first model
calculation of the full phase diagram at nonzero baryon and isospin density,
thus matching the model-independent predictions of chiral perturbation theory at
low density to the conjectured phase structure at high density. We confirm the
presence of the Fulde--Ferrell (FF) phase in the phase diagram and study its
dependence on the tunable parameter in the Lagrangian that simulates the
effects of the quantum axial anomaly. As a byproduct, we clarify the calculation
of the thermodynamic potential in the presence of the FF pairing, which was
previously based on an \emph{ad hoc} subtraction of an unphysical cutoff
artifact. Furthermore, we argue that close to the diquark (or pion)
Bose--Einstein condensation transition, the system behaves as a dilute Bose gas
so that our simple fermionic model in the mean-field approximation is not
quantitatively adequate. We suggest that including thermal fluctuations of the
order parameter for Bose--Einstein condensation is crucial for understanding
available lattice data.
\end{abstract}


\maketitle


\section{Introduction}
The attempts to understand the phase structure of quantum chromodynamics (QCD)
together with the infamous sign problem at nonzero baryon chemical potential,
have triggered interest in several QCD-like theories. These include QCD at
nonzero isospin density \cite{Son:2000xc}, QCD with adjoint quarks and two-color
QCD \cite{Kogut:1999iv,*Kogut:2000ek}. The common feature of these theories is
that they are free of the sign problem. This on the one hand admits their
straightforward simulation using lattice Monte-Carlo techniques and a check of
these methods against model-independent predictions of chiral perturbation
theory. On the other hand, it allows a direct test of various model approaches
at high temperature and/or density where chiral perturbation theory is not
applicable. Hence the study of QCD-like theories contributes to our
understanding of the physics of strongly-coupled gauge theories at nonzero
temperature and density.

In this paper we investigate the thermodynamics of two-color QCD, which has
attracted most interest among the QCD-like theories. We focus on the case of
two light quark flavors with nonzero baryon and isospin chemical potentials.
Although the underlying mechanisms of confinement and chiral symmetry breaking
in two-color QCD are believed to be the same as in the real, three-color QCD,
the actual physical behavior is quite different, especially at nonzero baryon
density. First, two quarks can combine into a color singlet (leaving the
gauge symmetry unbroken) which implies that at high density where Cooper pairing
near the Fermi surface occurs, the system behaves as a baryon superfluid rather
than a color superconductor. Second, even in the confined phase, the colorless
baryons are formed by two quarks, which means that ``nuclear'' matter is
actually a Bose gas and finite baryon density is realized through Bose--Einstein
condensation (BEC).

The very fact that two-color QCD is free of the sign problem follows from the
(pseudo)reality of the gauge group representation in which the quarks
transform. This has another notable consequence: the Pauli--G\"ursey symmetry
\cite{Smilga:1994tb,Rapp:1997zu} connecting quarks with antiquarks. As a
result, the spectrum of Nambu--Goldstone (NG) bosons of the spontaneously
broken global flavor symmetry includes pseudoscalar mesons as well as diquarks
in a single multiplet. In fact, it is the presence of light particles carrying
baryon number that allows to study the phase diagram of two-color QCD at nonzero
temperature and baryon density using the techniques of effective
field theory
\cite{Kogut:2000ek,Splittorff:2000mm,Splittorff:2001fy,*Splittorff:2002xn}.

While chiral perturbation theory provides a model-independent parameterization
of all observables in terms of a few low-energy couplings, its validity is
limited to the region of low density. Several approaches have therefore
been used to extend its results. These include the linear sigma model
\cite{Wirstam:1999ds,*Lenaghan:2001sd,*Wirstam:2002be}, random matrix
theory \cite{Vanderheyden:2001gx,*Klein:2004hv,*Kanazawa:2009en}, or 
high-density effective theory \cite{Kanazawa:2009ks}. The Nambu--Jona-Lasinio
(NJL) model was first applied to two-color QCD in Ref.~\cite{Kondratyuk:1991hf}
for one quark flavor and in Ref.~\cite{Ratti:2004ra} for two flavors. In
\cite{Sun:2007fc} the crossover from BEC of molecular diquarks to the
Bardeen--Cooper--Schrieffer (BCS) pairing of quarks was studied. Nonzero isospin
chemical potential imposes stress on this pairing and when sufficiently large,
induces a first-order transition to the normal phase. Around this transition, a
narrow window with inhomogeneous pairing may exist, as conjectured in
\cite{Splittorff:2000mm} and explicitly verified in \cite{Fukushima:2007bj}.
Despite the number of papers on the subject pursuing different analytic
approaches, the main body of work has been done on the lattice, using both the
strong-coupling expansion
\cite{Polonyi:1982wz,*Nishida:2003uj,*Chandrasekharan:2006tz,*Fukushima:2008su}
and Monte-Carlo simulations
\cite{Hands:2000ei,*Kogut:2001na,*Hands:2001ee,*Kogut:2003ju,*Skullerud:2003yc,%
*D'Elia:2006xb,Hands:2005yq,*Hands:2006ve,*Hands:2010gd}.

The plan of the paper is as follows. In Sec.~\ref{Sec:Lagrangians} we introduce
the NJL model for two-color QCD and demonstrate how it can be derived in a way
which makes the extended symmetry of two-color QCD manifest. Section
\ref{Sec:MFapprox} is entirely devoted to the mean-field approximation within
this model. The reason is that inclusion of the Fulde--Ferrell (FF) pairing
brings nontrivial regularization issues, in particular in connection with the
widely used sharp three-dimensional cutoff. Section \ref{Sec:phase_diagram}
presents the numerical results for the phase diagram at nonzero baryon and
isospin density. We include chiral, diquark and pion condensates with the
possibility of spatial (FF) modulation of the latter two. Using a generalized
NJL interaction, we also allow for isospin breaking by different chiral
condensates of the $u$ and $d$ quarks. We confirm the presence of the FF phase
at high chemical potential, in agreement with \cite{Fukushima:2007bj}. We show
that the region in the phase diagram where FF pairing occurs depends sensitively
on the parameter in the Lagrangian that measures axial symmetry breaking by the
quantum anomaly. In Sec.~\ref{Sec:collective} we investigate the collective
modes in the NJL model. We calculate their propagators and determine their
masses and dissociation temperatures. In Sec.~\ref{Sec:BEC} we discuss in some
depth two-color quark matter at nonzero baryon and zero isospin density, which
corresponds to the conditions being simulated on the lattice. Recent progress in
the lattice computations led to first attempts to probe BEC of diquarks and the
region of moderate baryon density
\cite{Hands:2005yq,*Hands:2006ve,*Hands:2010gd}. One of the motivations for the
present work was an attempt to understand the available lattice data on the
thermodynamics of cold and dense two-color quark matter. Even though we have not
been able to demonstrate quantitative agreement between the lattice simulation
and model calculations, we outline a possible direction along which it can
presumably be reached. Finally, in Sec.~\ref{Sec:conclusions} we summarize and
conclude.


\section{NJL Lagrangians for two-color quark matter}
\label{Sec:Lagrangians}
In any model attempting to mimic some features of full QCD, it is
mandatory that its Lagrangian has the same symmetries as the underlying
theory. In \cite{Ratti:2004ra} the NJL interaction for two-color QCD was derived
by a Fierz transformation of the four-quark interaction inspired by one-gluon
exchange, in which the flavor symmetry is manifest. Here we introduce a
formalism which makes it possible to construct NJL interactions with desired
symmetry properties directly. Our discussion will parallel that of Kogut
\emph{et al.}~\cite{Kogut:2000ek}. Yet we believe it is worthwhile to show some
details, since instead of working with the Weyl spinors for the quark fields as
in \cite{Kogut:2000ek}, we keep the Dirac spinors and introduce the Nambu
doublet notation, as is common in the literature on the NJL model.

Our starting point is the Lagrangian for the quark sector of two-color QCD,
\begin{equation}
\La_{\text{2cQCD}}=\adj\psi i\slashed D\psi-m_0\adj\psi\psi,
\label{QCD_Lagrangian}
\end{equation}
where $D_\mu\psi=(\de_\mu-ig\sigma_aA^a_\mu)\psi$ is the color-$\gr{SU(2)}$
covariant derivative, $m_0$ is the current mass common to all quark flavors, and
the flavor indices are suppressed. Next we replace the usual Dirac spinor,
consisting of the left- and right-handed components, with the purely left-handed
Nambu spinor and its (conveniently defined) charge conjugate,
$$
\Psi=\begin{pmatrix}
\psi_L\\
\sigma_2\cc{\psi_R}
\end{pmatrix},\quad
\cc\Psi=\sigma_2\begin{pmatrix}
\cc{\psi_L}\\
\cc{(\sigma_2\cc{\psi_R})}
\end{pmatrix}=
\begin{pmatrix}
\sigma_2\cc{\psi_L}\\
-\psi_R
\end{pmatrix},
$$
where $\cc\psi=C\adj\psi^T$ is usual Dirac charge conjugation. The two-color QCD
Lagrangian (\ref{QCD_Lagrangian}) can then be rewritten in the block-matrix form
\begin{equation}
\La_{\text{2cQCD}}=\adj\Psi i\slashed D\Psi-\left[
\frac12m_0\adj{\cc\Psi}
\begin{pmatrix}
0 & \openone\\
-\openone & 0
\end{pmatrix}
\Psi+\text{H.c.}\right].
\label{QCD_Nambu}
\end{equation}
All entries of the $2\times2$ Nambu mass matrix are themselves matrices in 
flavor space. Without going into details which are covered in literature
\cite{Kogut:2000ek}, we note that thanks to the (pseudo)reality of the
gauge-$\gr{SU(2)}$ generators, $\La_{\text{2cQCD}}$ with $N_f$ quark flavors
has an apparent $\gr{SU}(2N_f)$ symmetry in the chiral limit. The baryon
number, being generated by
$$
B=\frac12\begin{pmatrix}
\openone & 0\\
0 & -\openone
\end{pmatrix},
$$
is incorporated as a part of this simple group, while a mere change of the phase
of $\Psi$ corresponds to axial $\gr{U(1)_A}$ transformations. This symmetry is
known to be broken at the quantum level by the axial anomaly.

As is clear from Eq.~(\ref{QCD_Nambu}), the mass operator in the Nambu notation
is proportional to
$$
\adj{\cc\Psi}\Sigma_1\Psi=\Psi^TC\sigma_2\Sigma_1\Psi,\quad
\Sigma_1=
\begin{pmatrix}
0 & -\openone\\
\openone & 0
\end{pmatrix}.
$$
The standard chiral condensate which represents the ground state in the vacuum
has the same flavor structure as the mass term. Therefore, the multiplet of
mesons, one of which is to condense in the ground state and the others to become
NG modes, transform as a rank-2 antisymmetric tensor of $\gr{SU}(2N_f)$.
For $N_f=2$ we make use of the local group isomorphism $\gr{SU(4)\simeq SO(6)}$
to conclude that with a suitable basis of $4\times4$ antisymmetric matrices,
$\vec\Sigma=\{\Sigma_i\}_{i=1}^6$, the operator $\adj{\cc\Psi}\vec\Sigma\Psi$
transforms as a complex vector of $\gr{SO(6)}$. The basis matrices are unitary
and satisfy the orthogonality condition
$\Sigma_i\he{\Sigma_j}+\Sigma_j\he{\Sigma_i}=2\delta_{ij}$. They may be chosen
to carry the quantum numbers of the individual meson modes: three pions, sigma,
diquark and the antidiquark. The explicit form of the matrices with these
properties was given in \cite{Brauner:2006dv}.

It is now obvious how to construct a NJL interaction Lagrangian with 
$\gr{SO(6)}$ symmetry. As long as we only consider the Lorentz-scalar bilinear
operator $\adj{\cc\Psi}\vec\Sigma\Psi$, thus selecting bosonic modes with
specific quantum numbers \footnote{The particle content of our model Lagrangian
is minimal compatible with the symmetry of two-color QCD. In contains only the
modes which can be obtained from the ground state by a symmetry transformation,
and are guaranteed to exist by the Goldstone theorem.}, there are only two ways
to make an $\gr{SO(6)}$ invariant out of the vector
$\adj{\cc\Psi}\vec\Sigma\Psi$: taking a scalar product with
itself or its complex conjugate,
\begin{equation}
\begin{split}
\La_{\gr{U(4)}}&=G\bigl|\adj{\cc\Psi}\vec\Sigma\Psi\bigr|^2,\\
\La_{\gr{SU(4)}}&=\frac G2
\left[\bigl(\adj{\cc\Psi}\vec\Sigma\Psi\bigr)^2+\text{H.c.}\right].
\end{split}
\label{Nambu_Lagrangians}
\end{equation}
The subscripts emphasize the important fact that while $\La_{\gr{U(4)}}$
is invariant under the $\gr{U(1)_A}$ group of phase transformations of $\Psi$,
$\La_{\gr{SU(4)}}$ obviously breaks it. Inserting explicit expressions for
the $\Sigma_i$ matrices, these two interactions can be readily rewritten in
terms of the original Dirac spinor $\psi$. It is convenient to work with the
following combinations,
\begin{align*}
\La_1&=\frac12\left[\La_{\gr{U(4)}}+\La_{\gr{SU(4)}}\right]\\
&=G\left[(\adj\psi\psi)^2+(\adj\psi i\gamma_5\vec\tau\psi)^2
+|\adj{\cc\psi}\gamma_5\sigma_2\tau_2\psi|^2\right],\\
\La_2&=\frac12\left[\La_{\gr{U(4)}}-\La_{\gr{SU(4)}}\right]\\
&=G\left[(\adj\psi i\gamma_5\psi)^2+(\adj\psi\vec\tau\psi)^2
+|\adj{\cc\psi}\sigma_2\tau_2\psi|^2\right],
\end{align*}
where $\vec\tau$ are Pauli matrices in isospin space. The quantum numbers of
the various channels are now apparent: $\La_1$ contains the sigma, pions and the
scalar (anti)diquark, while $\La_2$ contains their parity partners.

Note that $\La_1$ is essentially the minimal two-flavor NJL Lagrangian
\cite{Klevansky:1992qe}, augmented by a diquark term. The only difference with
Lagrangians used in model investigations of dense (three-color) quark matter is
that here the $\gr{SU(4)}$ symmetry fixes the coupling in the diquark channel
to be the same as for the meson terms. This is the Lagrangian used in
\cite{Ratti:2004ra,Sun:2007fc}. However, since we have two independent
interaction terms in Eq.~(\ref{Nambu_Lagrangians}), the most general
interaction one can construct is a linear combination
\begin{equation}
\La_{\text{int}}=\La_1+\zeta\La_2,
\label{L_int}
\end{equation}
and was written down for the first time in \cite{Brauner:2009gu}. At $\zeta=1$
the Lagrangian reduces to $\La_{\gr{U(4)}}$, preserving the axial $\gr{U(1)_A}$,
and is a direct descendant of the current--current interaction used in
\cite{Ratti:2004ra}. The case of $\zeta=-1$, that is, $\La_{\gr{SU(4)}}$,
corresponds to the two-flavor instanton-induced interaction, breaking 
$\gr{U(1)_A}$ maximally \cite{Rapp:1997zu}. In the present paper $\zeta$ is
treated as a free parameter, the only restriction being $\zeta\leq1$. This
follows from the fact that in the vacuum, the scalar chiral condensate must be
preferred over the pseudoscalar one, as guaranteed by the  Vafa--Witten theorem
\cite{Vafa:1984xg}. We note in passing that in the three-color NJL model, one
usually introduces the parameter $\alpha=(1-\zeta)/2$ which measures explicit
$\gr{U(1)_A}$ breaking by the axial anomaly so that the interaction Lagrangian
can also be expressed as
$\La_{\text{int}}=(1-\alpha)\La_{\gr{U(4)}}+\alpha\La_{\gr{SU(4)}}$. The
typical values of $\alpha$ lie somewhere in the range $0.1-0.2$
\cite{Buballa:2003qv}.

The full NJL Lagrangian now reads
$$
\La=\adj\psi\bigl[i\slashed\de-m_0+\tfrac12\gamma_0(\mu_B+\tau_3\mu_I)\bigr]
\psi+\La_{\text{int}},
$$
where $\mu_B$ and $\mu_I$ are baryon number and isospin chemical potentials,
both being assumed positive without lack of generality. In \cite{Brauner:2009gu}
this Lagrangian with the generalized interaction (\ref{L_int}) was employed to
study the masses of the scalar mesons and hence the axial symmetry restoration
in the meson spectrum. Here it will be crucial at high baryon and isospin
chemical potentials, for it allows to introduce different $u$-quark and
$d$-quark chiral condensates within the mean-field approximation.


\section{Mean-field approximation}
\label{Sec:MFapprox}
As mentioned above, the scalar chiral and diquark condensates are preferred
to the pseudoscalar ones. Since we will not be interested in the corresponding
non-NG modes (eta meson and the pseudoscalar diquark) either, we will
completely neglect the interaction terms in the $\adj\psi i\gamma_5\psi$ and
$\adj{\cc\psi}\sigma_2\tau_2\psi$ channels. The remaining four channels can be
rewritten using the Hubbard--Stratonovich transformation. The Lagrangian can
then be written as
\begin{multline}
\La=\adj\psi\bigl[
i\slashed\de-m_0+\tfrac12\gamma_0(\mu_B+\tau_3\mu_I)-\sigma-
i\gamma_5\vec\tau\cdot\vec\pi-\zeta\vec\tau\cdot
\vec\rho\bigr]\psi\\
+\frac12(\Delta^*\adj{\cc\psi}\gamma_5\sigma_2\tau_2\psi+\text{H.c.})
-\frac1{4G}(\sigma^2+\vec\pi^2+\zeta\vec\rho^2+|\Delta|^2),
\label{Lagr_bosonized}
\end{multline}
where $\sigma,\vec\pi,\vec\rho,\Delta$ are the collective fields with the
quantum numbers of a scalar isoscalar, pseudoscalar isovector, scalar
isovector, and scalar diquark, respectively.

In the mean-field approximation the collective fields are replaced with their
vacuum expectation values. The presence of both baryon number and isospin
chemical potentials induces a mismatch between the Fermi levels of quarks of
different flavors. When the mismatch is too large, it breaks the BCS-like
pairing and the system goes over to the normal phase via a first-order phase
transition. Around this transition, a narrow window may exist where pairing
with nonzero total momentum occurs. Such kind of pairing, well known in
condensed matter physics, was first considered in the context of dense quark
matter in \cite{Alford:2000ze,*Bowers:2001ip}, see also
\cite{Casalbuoni:2003wh} for a comprehensive review.

In literature on color superconductivity of dense (three-color) quark
matter (see \cite{Alford:2007xm,*Wang:2009xf} for recent reviews), there are
in fact several other candidates for the state of dense matter in presence of a
Fermi surface mismatch that compete with the inhomogeneous (FF) pairing. These
include phases with a gluon condensate \cite{Gorbar:2005rx,*Gorbar:2006up},
meson condensate \cite{Bedaque:2001je,*Kaplan:2001qk}, or a meson supercurrent
\cite{Schaefer:2005ym,*Kryjevski:2005qq}. Let us explain briefly why we do not
consider these possibilities in the present context. First, in a two-flavor
(three-color) color superconductor, the single plane wave FF state is equivalent
to the BCS state with a uniform gluon condensate; they are connected by a gauge
transformation \cite{Gorbar:2005rx,*Gorbar:2006up}. However, this is not the
case in two-color QCD where the diquark pairs are color singlets. For the same
reason, the gluon condensate will be disfavored since it induces nonzero color
charge (which in three-color QCD compensates for the charge of the Cooper
pairs).

The possibility of a meson condensate strongly depends on the meson
spectrum, and hence on the pattern of global symmetry breaking. In three-color
QCD meson condensation is likely to occur in three-flavor matter where there is
an octet of light pseudoscalar mesons \cite{Bedaque:2001je,*Kaplan:2001qk}.
On the other hand, in two-color QCD with two flavors, considered in this paper,
the only mesons present in the low-energy spectrum of the diquark condensation
phase are the pions. It can be shown that at $\mu_I=0$, the pion mass is equal
to $\mu_B$ \cite{Ratti:2004ra}. This eventually does lead to pion condensation,
but only when $\mu_I>\mu_B$. There is no meson to condense \emph{inside} the
diquark condensation phase, thus leaving the FF pairing the only candidate for
resolving the Fermi surface mismatch.

In the present paper, we will therefore take into account the possibility of
inhomogeneous diquark and pion condensates. To that end, note that the
Lagrangian of two-color QCD (\ref{QCD_Lagrangian}) is invariant (among
others) under the discrete symmetry of charge conjugation of $d$-quarks. The
reason why we can charge conjugate one component of the quark flavor doublet and
still keep the Lagrangian gauge invariant is once again the (pseudo)reality of
the doublet representation of the gauge-$\gr{SU(2)}$. This discrete symmetry
remains intact even in presence of the baryon number and isospin chemical
potentials, provided they are interchanged simultaneously with the charge
conjugation. As a consequence, the phase diagram of two-color QCD in the
$(\mu_B,\mu_I)$ plane is symmetric with respect to  interchanging of the axes
accompanied by the simultaneous interchanging of the diquark and pion
condensates \footnote{The fact that pions, unlike the diquark, form an isospin
triplet is immaterial. In fact, any nonzero value of $\mu_I$ breaks the isospin
symmetry explicitly to its $\gr{U(1)}_{I_3}$ subgroup generated by the third
component of isospin, which the chemical potential is associated with. The
charged pions then carry one unit of this unbroken charge, very much like the
diquark carries baryon number.}. Moreover, the diquark and pion condensates
never occur simultaneously, the former being favored for $\mu_B>\mu_I$, while
the latter for $\mu_B<\mu_I$. This is rather obvious at high chemical potentials
where quarks form a Fermi sea: at
$\mu_B>\mu_I$ a Fermi sea of $d$-quarks exists, thus making formation of $ud$
pairs, and hence diquark condensation, possible. Likewise, at $\mu_B<\mu_I$,
$\bar d$-quarks form a Fermi sea, leading to $u\bar d$ Cooper pairs and pion
condensation. At low chemical potentials, the mutual exclusion of the pion and
diquark condensates may be verified using chiral perturbation theory.

With the above argument in mind, we will set $\vec\pi=\vec0$ and consider
the region $0<\mu_I<\mu_B$. The full phase diagram is easily constructed
by reflection with respect to the line $\mu_B=\mu_I$, accompanied by the
replacement of the diquark condensate with the pion one. For the coordinate
dependence of the diquark condensate we will assume the simplest, single
plane wave form, $\Delta e^{2i\vek q\cdot\vek x}$ with real $\Delta$. This is
the FF pairing \cite{Fulde:1964ff}. As a side remark let us add that
the FF pairing is not really the energetically most favored state. The energy
may be further decreased by forming a more complicated spatial structure such as
the truly inhomogeneous standing wave-like Larkin--Ovchinnikov state
\cite{Larkin:1964zz}, or even various crystalline structures
\cite{Bowers:2002xr,*Rajagopal:2006ig}. When the periodic structure of the order
parameter is not limited to a linear combination of a finite number of plane
waves, the first order transition between the BCS pairing and the FF phase
is weakened and eventually becomes second order, enlarging the region in the
phase diagram occupied by the inhomogeneous phase
\cite{Matsuo:1998ma,*Nickel:2008ng}. FF pairing in the context of pion
condensation was investigated before in \cite{He:2006tn,*Huang:2007cr}.

All in all, we take into account the following mean fields: $\sigma$,
$\rho$ (standing for the third component of $\vec\rho$), and $\Delta
e^{2i\vek q\cdot\vek x}$. Note that $\rho$ gives rise to splitting of the
constituent quark masses, being $M_{u,d}=m_0+\sigma\pm\zeta\rho$. The coordinate
dependence of the diquark condensate may be removed by a unitary transformation
at the price of modifying the kinetic term of quarks. After integrating out the
quarks and performing the Matsubara summation, the thermodynamic potential
density of the model becomes
\begin{equation}
\begin{split}
\frac{\Omega(\sigma,\rho,\Delta,q)}V=
&\frac1{4G}(\sigma^2+\zeta\rho^2+\Delta^2)\\
&-T\int\frac{d^3\vek k}{(2\pi)^3}
\tr\log\left(1+e^{-\beta\Ha_{\vek k}}\right).
\label{TDpot}
\end{split}
\end{equation}
Here $\Ha_{\vek k}$ is the $16\times16$ matrix Hamiltonian whose block
elements in the Nambu space $(\psi_r,\sigma_2\cc{\psi_g})$, read
\begin{align*}
(\Ha_{\vek k})_{11}=&\vek\alpha\cdot(\vek k+\vek q)
+\gamma_0(M+\zeta\rho\tau_3)
-\frac12(\mu_B+\tau_3\mu_I),\\
(\Ha_{\vek k})_{12}=&-\gamma_0\gamma_5\Delta,\\
(\Ha_{\vek k})_{21}=&+\gamma_0\gamma_5\Delta,\\
(\Ha_{\vek k})_{22}=&\vek\alpha\cdot(\vek k-\vek q)
+\gamma_0(M-\zeta\rho\tau_3)
+\frac12(\mu_B-\tau_3\mu_I),
\end{align*}
where we introduced the notation $M=m_0+\sigma$ for the average constituent
quark mass, and $\vek\alpha=\gamma_0\vek\gamma$ are the usual Dirac matrices.
Note that the zero-point energy contribution is missing in Eq.~(\ref{TDpot})
since $\tr\Ha_{\vek k}=0$. The eigenvalues of the Hamiltonian, representing the
energies of fermionic quasiparticles, cannot be found analytically. When
calculating the thermodynamic potential, one has to resort to numerical
diagonalization of the Hamiltonian. In \cite{Fukushima:2007bj} this technical
difficulty was avoided by using an approximate analytic expression for the
quasiparticle dispersion relations. In the present paper the thermodynamic
potential is evaluated without introducing any further approximation.

The naive expression for the thermodynamic potential (\ref{TDpot}) is actually
ill-defined because the momentum integral is badly divergent. The integral will
be regulated with a three-momentum cutoff, $\Lambda$, as usual. However, with
nonzero FF momentum $\vek q$ one must be very careful at what stage of the
calculation the cutoff is applied. Trying to impose it naively to
Eq.~(\ref{TDpot}) leads to unphysical artifacts including a $\vek q$-dependent
quadratic divergence. This was previously removed by an \emph{ad hoc}
subtraction \cite{Fukushima:2007bj,Gorbar:2005tx}, making sure that the
thermodynamic potential is independent of $\vek q$ once $\Delta=0$, which
follows from the very definition of the FF order parameter. We would like to
argue here that such a subtraction follows naturally when the thermodynamic
potential is first properly regularized, and only after then the unitary
transformation leading to Eq.~(\ref{TDpot}) is performed. Since the argument is
rather general and applies in particular also to the analogous nonrelativistic
problem, we present it separately in the Appendix. The final expression for the
subtracted thermodynamic potential, using the function
$\Omega(\sigma,\rho,\Delta,q)$ defined by Eq.~(\ref{TDpot}), is
\begin{equation}
\begin{split}
\Omega_{\text{sub}}(\sigma,\rho,\Delta,q)=&
\Omega(\sigma,\rho,\Delta,q)\\
&-\Omega(\sigma,\rho,0,q)+\Omega(\sigma,\rho,0,0).
\end{split}
\label{subtr_Omega}
\end{equation}
The addition and subtraction operations are understood to be performed inside
the integration. This prescription, differing slightly from that used in
\cite{Fukushima:2007bj}, also ensures that the thermodynamic
potential is independent of $\vek q$ once $\Delta=0$. Moreover, the
thermodynamic potential $\Omega_{\text{sub}}$ has a well defined minimum which
determines the equilibrium state.


\section{Phase diagram at nonzero baryon and isospin density}
\label{Sec:phase_diagram}
The phase diagram of two-color QCD with two quark flavors in the
$(\mu_B,\mu_I)$ plane and at zero temperature was first drawn in
\cite{Splittorff:2000mm}. At low chemical potential its structure can be
determined in a model-independent manner using the chiral perturbation theory.
One thus finds that when both chemical potentials are smaller than $m_\pi$, the
vacuum mass of the pion--diquark multiplet of pseudo-NG bosons of the
spontaneously broken $\gr{SU(4)}$ flavor symmetry, the system is in the vacuum
phase. Only the chiral condensate is nonzero, equal to its vacuum value
independently of the chemical potential. As soon as one of the chemical
potentials, say $\mu_B$ (remember the reflection symmetry of the phase
diagram), exceeds $m_\pi$, the mode carrying the associated charge---in this
case the diquark---condenses via a second-order phase transition. Since this
picture relies just on the symmetry and the pattern of its spontaneous
breaking, it is not surprising that results obtained within chiral perturbation
theory, lattice simulations \cite{Hands:2000ei,*Hands:2001ee}, and model
calculations \cite{Ratti:2004ra} are in good quantitative agreement. At
moderate chemical potentials one therefore finds BEC phases with a diquark (for
$\mu_B>\mu_I$) or pion condensate (for $\mu_B<\mu_I$), separated by a
first-order phase transition line at $\mu_B=\mu_I$.

On the other hand, at high chemical potential one expects the physics of the
system to be dominated by the Fermi sea of quarks. Spontaneous symmetry
breaking is realized by a formation of Cooper pairs via the standard BCS
scenario. Since the Cooper pairs carry the same quantum numbers as the
condensing modes in the low-density BEC phase, there is no phase transition
between the two regimes despite the fact that they feature qualitatively
different physical behavior. One speaks of the BCS--BEC crossover
\cite{Eagles:1969ea,*Leggett:1980le}. In the context of two-color QCD, this was
studied in \cite{Sun:2007fc}; we will comment on it later in
Sections \ref{Sec:collective} and \ref{Sec:BEC}.

The qualitative shape of the phase diagram at high chemical potential was
conjectured in \cite{Splittorff:2000mm} based on the experience with real,
three-color QCD at weak coupling. The essence of the argument is as follows.
Close to the $\mu_B$ and $\mu_I$ axes, one will find diquark and pion
condensates as predicted by BCS theory. However, since both of them involve
pairing of quarks of different flavors, they will become less and less favored
towards the ``diagonal'' of the phase diagram; $\mu_I$ generates a mismatch
between the Fermi levels of $u$ and $d$ quarks in the diquark condensation
phase, while $\mu_B$ plays the same role in the pion condensation phase. When
the mismatch becomes large enough, the BCS pairing is no longer energetically
favorable and it breaks in a first-order phase transition. In a narrow window
around this transition, a phase where Cooper pairs carry nonzero total momentum
may exist. Even after the cross-flavor pairing is made impossible, quarks
of the same flavor can still pair, since there is no Fermi level mismatch by
definition. Such pairs are required to carry nonzero spin or orbital momentum by
the Pauli principle \cite{Kondratyuk:1991hf}. In a narrow strip along the
diagonal of the phase diagram, the chemical potential of $d$-quarks is almost
zero, and they should therefore be confined and exhibit the one-flavor chiral
condensate. Thus, only the $u$-quarks will pair in this region.

\begin{figure}
\begin{center}
\includegraphics[scale=1]{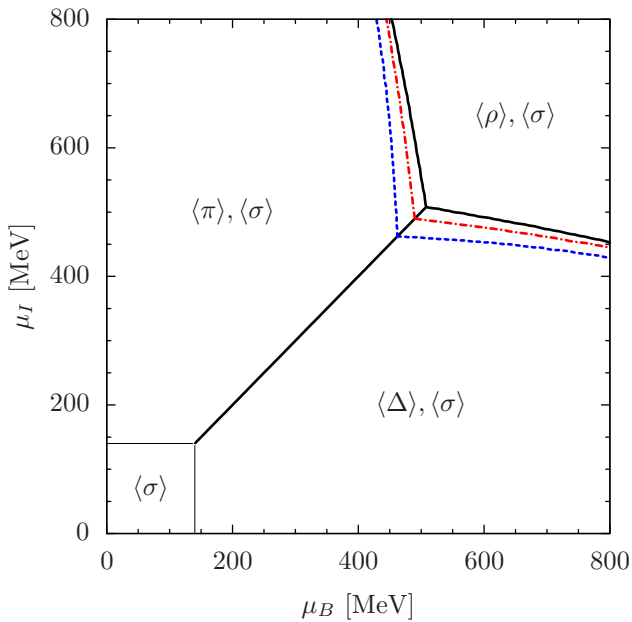}
\end{center}
\caption{Phase diagram of two-color QCD with two quark flavors in the
$(\mu_B,\mu_I)$ plane at zero temperature. Various phases are labeled by the
corresponding nonzero condensates. The $\rho$ condensate in the upper right
corner only appears at $\zeta\neq0$. Thick and thin lines denote first- and
second-order transitions, respectively. The black solid lines were
obtained with $\zeta=0$ ($\alpha=0.5$), the red dash-dotted lines with
$\zeta=0.8$ ($\alpha=0.1$), and the blue dashed lines with $\zeta=1$
($\alpha=0$). The position of the transitions at high chemical potentials only
changes appreciably for $\zeta$ close to one. The $\zeta=0.6$ lines would almost
coincide with the $\zeta=0$ ones shown.}
\label{Fig:PDfull}
\end{figure}
We employed the NJL model to verify the conjecture made in
\cite{Splittorff:2000mm}. In order to obtain concrete numbers, we used the
two-color NJL parameter set, established in \cite{Brauner:2009gu}:
$G=7.23\text{ GeV}^{-2}$, $m_0=5.4\text{ MeV}$, and $\Lambda=657\text{ MeV}$.
This corresponds to the physical pion mass $m_\pi=140\text{ MeV}$, pion decay
constant $f_\pi=75.4\text{ MeV}$, and (one-flavor) chiral condensate
$\langle\adj\psi_u\psi_u\rangle=(-218\text{ MeV})^3$. The latter two quantities
were obtained from usual three-color values by rescaling with respect to the
number of colors. The parameter $\zeta$ is treated as free in this paper.

While we included the FF-type diquark and pion condensates as well as
independent chiral condensates for $u$ and $d$ quarks, we did not, for the sake
of simplicity, take into account spin-one pairing. This would most likely
further split the region in the upper right corner of the phase diagram into
domains with or without spin-one pairing of individual flavors. On the other
hand, it is not expected to alter the structure that we find. The reason is
that spin-one pairing is usually characterized by a small pairing gap as
compared to spin-zero pairing, and the pairing moreover does not occur
isotropically on the whole Fermi surface. Consequently, the energy gain from
spin-one pairing is minute so that it cannot compete with phases with other
types of pairing.

\begin{figure}
\begin{center}
\includegraphics[scale=1]{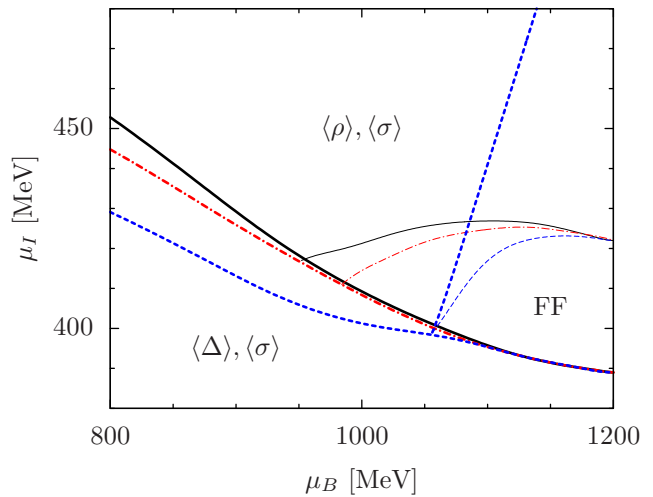}
\end{center}
\caption{The high chemical potential corner of the phase diagram. The notation
for the lines is the same as in Fig.~\ref{Fig:PDfull}. The region where the
momentum $\vek q$ is nonzero is indicated as ``FF''. The additional thick blue
dashed line is associated with the first-order chiral restoration transition
at $\xi=1$,  explained in the text. Note that the chemical potential used to
label the horizontal axis is the baryon number one so that the corresponding
quark chemical potential in the range displayed is still below the cutoff
$\Lambda$. Hence, cutoff artifacts, though non-negligible, will not alter the
qualitative structure of the phase diagram.}
\label{Fig:PDdetail}
\end{figure}
The  phase diagram at zero temperature obtained with the above set of parameters
and condensates taken into account is shown in Fig.~\ref{Fig:PDfull}. One can
see that at moderate chemical potentials, there is no FF phase and the system
goes directly to the phase with just chiral condensate(s). We avoid calling this
a normal phase, because the size of the chiral condensate(s) depends sensitively
on the parameter $\zeta$. For $\zeta=0$, $\rho=0$ and the common chiral
condensate of the $u$ and $d$ quarks is indeed very small. As $\zeta$ increases,
the flavor chiral condensates $\sigma_{u,d}=\sigma\pm\zeta\rho$ split, or in
other words, $\rho$ becomes nonzero. Naturally, $\sigma_u$ remains very small
since the $u$-quark chemical potential is very large, while $\sigma_d$ grows
towards the diagonal of the phase diagram. In fact, in the $\gr{U}(4)$ symmetric
limit ($\zeta=1$) the mean-field gap equations for $\sigma_u$ and $\sigma_d$ in
absence of diquark and pion condensates completely decouple. The $d$-quark
condensate $\sigma_d$ is then at the diagonal of the phase diagram (and around)
equal to its vacuum value. As soon as the $\gr{U(1)_A}$ symmetry is
explicitly broken, the flavors become entangled and the $d$-quark
condensate drops fast with increasing $\alpha$. However, despite the strong
dependence of the flavor chiral condensates on $\zeta$, the position of the
phase transition does not change much. Some shift is actually only visible as
$\zeta$ approaches one. Note that for the sake of clarity, we do not indicate
the BCS--BEC crossover in Fig.~\ref{Fig:PDfull}. This appears at
$\mu_{B,I}\approx229\text{ MeV}$; its position is to a very good precision
determined by the expression $(2M_0m_\pi^2)^{1/3}$, where $M_0$ is the
constituent quark mass in the vacuum \cite{Sun:2007fc}.

The domain of high chemical potentials is shown in detail in
Fig.~\ref{Fig:PDdetail}. The FF phase appears only at rather high chemical
potentials. This is in agreement with the observation that at in strongly
coupled systems the region with the FF phase is diminished
\cite{Fukushima:2007bj,Sedrakian:2009kb}. In quark matter one cannot tune the
coupling arbitrarily. Nevertheless, the weakly coupled BCS pairing is achieved
effectively at high chemical potential since the ratio $\Delta/\mu_B$ drops
with increasing chemical potential. Even though we used a different parameter
set than in \cite{Fukushima:2007bj} and a more refined method to evaluate the
thermodynamic potential at nonzero FF momentum $\vek q$, our results are in
reasonable quantitative agreement.

As $\zeta$ increases, the FF phase is obviously expelled from the phase diagram.
This is a direct consequence of the fact that the energy gain from FF pairing is
rather small due to both the small size of the pairing gap and the fact that
pairing occurs only on a small part of the Fermi surface. It therefore becomes
energetically more favorable to form a state with a large splitting of
constituent $u$ and $d$ quark masses, which in turn inhibits the cross-flavor
pairing. On the other hand, one should note that in the FF phase, $\rho$ does
take a nonzero, albeit small value. The reason is that the difference of the
constituent quark masses partially compensates for the chemical potential
mismatch, thus reducing the mismatch of Fermi \emph{momenta}.

Fig.~\ref{Fig:PDdetail} also features an additional first-order transition line
which occurs only for $\zeta=1$ and is defined by the condition
$\mu_B-\mu_I=\text{const}$. This can be explained using the observation made
above that at $\zeta=1$ the gap equations for $\sigma_u$ and $\sigma_d$
completely decouple. The two flavors then live their separate lives, and
instead of $\mu_{B,I}$ it would be more convenient to use the basis
$\mu_{u,d}=\frac12(\mu_B\pm\mu_I)$. The first-order line in question
corresponds to the critical value of $\mu_d$ at which a strong first-order
chiral symmetry breaking/restoration phase transition occurs. The critical
value, $\mu_d\approx330\text{ MeV}$, extracted from the phase diagram in
Fig.~\ref{Fig:PDdetail} , is in excellent agreement with an elementary
calculation within a one-flavor version of our model (\ref{Lagr_bosonized}),
keeping the chiral condensate $\sigma_d$ as the sole order parameter.

Once $\zeta$ drops below one, that is, axial $\gr{U(1)_A}$ symmetry is
explicitly broken by nonzero $\alpha$, the two quark flavors become entangled.
This smears the phase transition in the $d$-quark sector, making it a smooth
crossover. We have checked numerically that there is no sharp phase transition
already for very small $\alpha$, and for the case $\alpha=0.1$ displayed in
Fig.~\ref{Fig:PDdetail} by red dash-dotted lines, the crossover is already
rather gradual.
\begin{figure}
\begin{center}
\includegraphics[scale=1]{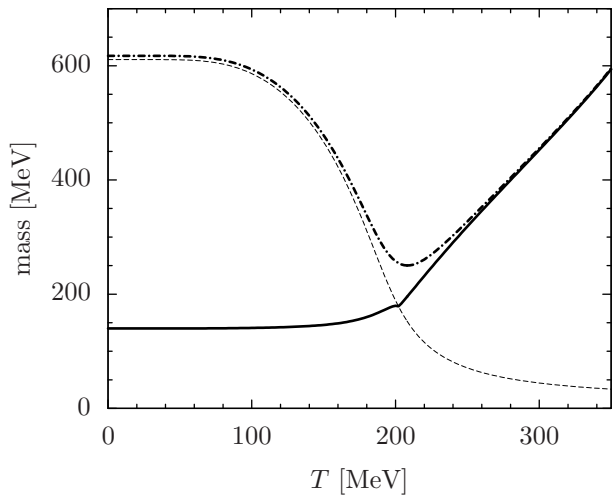}
\end{center}
\caption{Meson masses as a function of temperature at zero chemical potential.
Solid line: common mass of pions and the (anti)diquark. Dash-dotted line: sigma
mass. The thin dashed line indicates the position of the two-particle
continuum, that is, twice the constituent quark mass.}
\label{Fig:masses}
\end{figure}


\section{Collective modes}
\label{Sec:collective}
With the semi-bosonized Lagrangian (\ref{Lagr_bosonized}), the propagators of
the bosonic collective modes are straightforwardly derived by integrating out
the quarks and taking the second variational derivative with respect to the
collective fields $\chi_i$. The general expression for the inverse propagator
at imaginary (Matsubara) frequency $i\Omega_N$ is \cite{Brauner:2009gu}
\begin{multline*}
D_{ij}^{-1}(i\Omega_N,\vek p)=
\frac{1}{2G}\delta_{ij}+T\sum_n\int\frac{d^3\vek k}{(2\pi)^3}\\
\times\tr\left[\frac{\partial\Ha}{\partial\chi_i}
\frac{1}{i(\omega_n+\Omega_N)-\Ha_{\vek k+\frac{\vek p}{2}}}
\frac{\partial\Ha}{\partial\chi_j}
\frac{1}{i\omega_n-\Ha_{\vek k-\frac{\vek p}{2}}}
\right].
\end{multline*}
This formula is valid for the $\sigma,\vec\pi,\Delta$ modes we are
interested in. For modes from the other chiral multiplet, a factor $\zeta$ must
be added to the constant part of the propagator. In general, when some of the
condensates break the symmetry spontaneously, different modes may mix. We will
only give explicit expressions for the propagators in the case that just the
isospin-singlet chiral condensate ($\sigma$) is nonzero. This covers the
connected domain in the phase diagram containing the origin. Since the BEC
transition to either diquark or pion condensed phase is of second order, the
knowledge of these propagators allows one to locate these transitions by means
of the Thouless criterion \cite{Thouless:1960th}. For the sake of simplicity of
the resulting formulas, we will set the external three-momentum to zero. (A
more complete analysis of collective modes may be found in
\cite{Brauner:2009gu}; the evolution of the collective mode spectrum as a
function of baryon chemical potential has also been studied on the lattice
\cite{Hands:2007uc}.) With these simplifications in mind, the inverse
propagators of the six modes of interest to us read, after analytic continuation
to general complex frequency
$\omega$,
\begin{widetext}
\begin{equation}
\begin{split}
D^{-1}_{\sigma}(\omega)&=\frac{1}{2G}-4\sum_{e=\pm}
\int\frac{d^3\vek k}{(2\pi)^3}\frac{\vek k^2}{\eps^2_{\vek k}}
\left[\frac{1-f(e\xi^e_{\vek k})-f(e\xi^{-e}_{\vek k})}{\omega+2e\eps_{\vek k}}+
\frac{1-f(e\hat\xi^e_{\vek k})-f(e\hat\xi^{-e}_{\vek k})}{\omega+2e\eps_{\vek
k}}\right],\\
D^{-1}_{\pi^0}(\omega)&=\frac{1}{2G}-4\sum_{e=\pm}
\int\frac{d^3\vek k}{(2\pi)^3}
\left[\frac{1-f(e\xi^e_{\vek k})-f(e\xi^{-e}_{\vek k})}{\omega+2e\eps_{\vek k}}+
\frac{1-f(e\hat\xi^e_{\vek k})-f(e\hat\xi^{-e}_{\vek k})}{\omega+2e\eps_{\vek
k}}\right],\\
D^{-1}_{\pi^+}(\omega)&=\frac{1}{2G}-8\sum_{e=\pm}
\int\frac{d^3\vek k}{(2\pi)^3}
\frac{1-f(e\xi^e_{\vek k})-f(e\hat\xi^{-e}_{\vek k})}
{\omega+\mu_I+2e\eps_{\vek k}},\\
D^{-1}_{\Delta}(\omega)&=\frac{1}{2G}-8\sum_{e=\pm}
\int\frac{d^3\vek k}{(2\pi)^3}
\frac{1-f(e\xi^e_{\vek k})-f(e\hat\xi^{e}_{\vek k})}
{\omega+\mu_B+2e\eps_{\vek k}}.\\
\end{split}
\label{propagators}
\end{equation}
\end{widetext}
We use the usual notation for the Lorentz-invariant dispersion relation,
$\eps_{\vek k}=\sqrt{\vek k^2+M^2}$, and the Fermi--Dirac distribution function,
$f(x)=1/(e^{\beta x}+1)$. Furthermore, $\xi^e_{\vek k}=\eps_{\vek k}+\frac
e2(\mu_B+\mu_I)$ denotes the in-medium energy of the $u$-(anti)quark, and
$\hat\xi^e_{\vek k}=\eps_{\vek k}+\frac e2(\mu_B-\mu_I)$ that of the
$d$-(anti)quark. The propagators of $\pi^-$ and of the antidiquark are obtained
from those of $\pi^+$ and the diquark by changing the sign of $\mu_I$ and
$\mu_B$, respectively.

In Fig.~\ref{Fig:masses} we plot the meson masses as a function of temperature
at zero chemical potential. Note that thanks to the unbroken $\gr{SO(5)}$
symmetry, the pions and diquarks are exactly degenerate even away from the
chiral limit. The masses are determined from the zeroes of the real part of the
inverse propagators (\ref{propagators}). As long as the mode is stable, this
definition coincides with the physical, pole mass. Once the two-particle
threshold is crossed, the definition of the mass becomes ambiguous and it is
more convenient to discuss the full spectral density, $\rho(\omega)$. This is
obtained from the imaginary part of the retarded Green's function, and is
related to the inverse propagators (\ref{propagators}) by
$$
\rho(\omega)=-\frac{\mathrm{Im}\,D^{-1}(\omega+i\varepsilon)}
{[\mathrm{Re}\,D^{-1}(\omega+i\varepsilon)]^2+
[\mathrm{Im}\,D^{-1}(\omega+i\varepsilon)]^2}.
$$
\begin{figure}
\begin{center}
\includegraphics[scale=1]{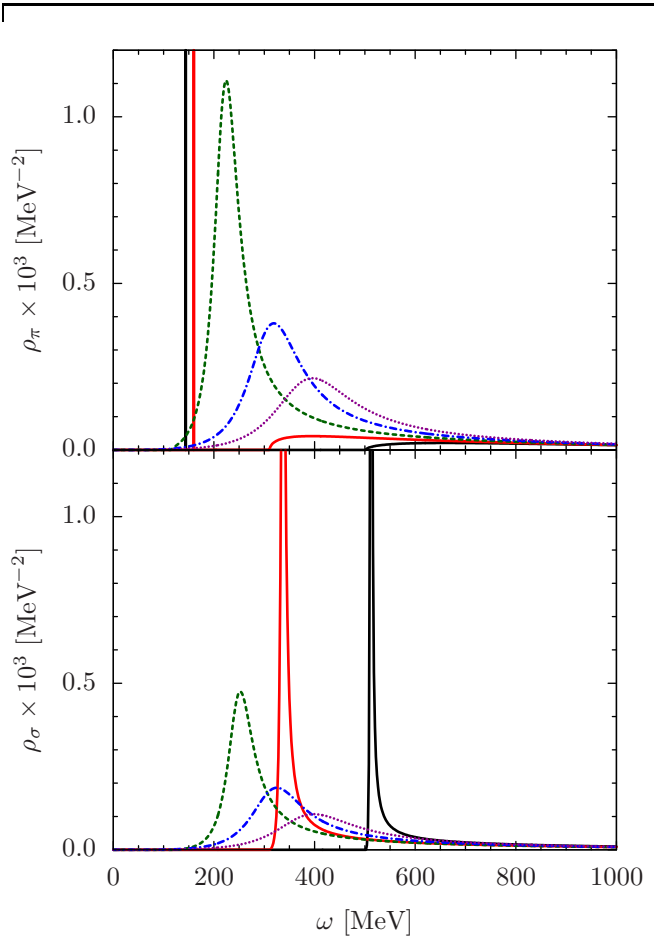}
\end{center}
\caption{Spectral densities of the pions and diquarks (upper panel) and sigma
(lower panel) as a function of frequency at zero chemical potential. The
various lines correspond to temperature of: $140\text{ MeV}$ (black, solid),
$180\text{ MeV}$ (red, solid), $220\text{ MeV}$ (green, dashed), $260\text{
MeV}$ (blue, dash-dotted), $300\text{ MeV}$ (magenta, dotted).}
\label{Fig:spectral}
\end{figure}%
The spectral densities as a function of frequency for five different values of
temperature ($140,180,220,260,300\text{ MeV}$) are plotted in
Fig.~\ref{Fig:spectral}. One can see clearly that all modes constitute
well-defined quasiparticles below the chiral restoration crossover; the pions
and diquarks are bound and therefore stable, while the sigma mass is just above
the threshold and thus gives rise to a very sharp peak in the spectrum. Once the
temperature increases, the constituent quark mass rapidly drops, the level of
the two-particle continuum consequently decreases, and all mesons become broad
and eventually more or less washed out of the spectrum. For temperatures above
about $300\text{ MeV}$ the notion of mass becomes meaningless.

\begin{figure}
\begin{center}
\includegraphics[scale=1]{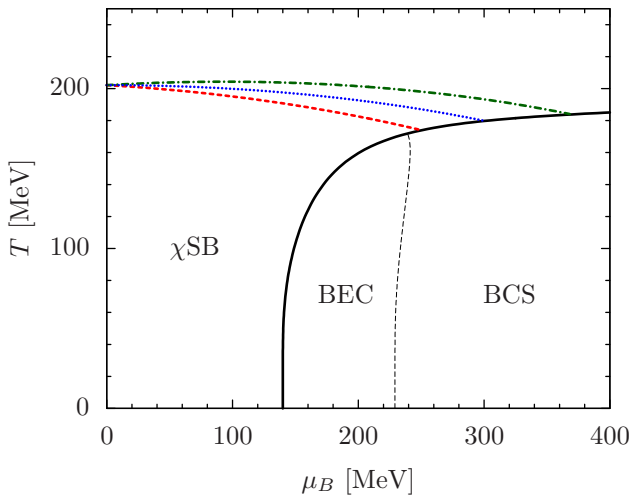}
\end{center}
\caption{Phase diagram of two-color QCD at zero isospin chemical potential. The
thick black solid line denotes the second-order diquark condensation
transition. The thin black dashed line indicates the position of the BCS--BEC
crossover, defined by the condition $M=\mu_B/2$. The dissociation temperatures
of the collective modes are represented by the: red dashed line for the
diquark, green dash-dotted line for the antidiquark, blue dotted line for the
pions. The region where the physics is dominated by the chiral condensate
is labeled as $\chi\text{SB}$.}
\label{Fig:phase_diagram}
\end{figure}
The range of stability of the modes is more precisely determined by the
so-called dissociation temperature, defined as the temperature at which the
quasiparticle pole in the propagator (\ref{propagators}) enters the two-particle
continuum \cite{Nishida:2005ds,*Abuki:2006dv,*Kitazawa:2007zs}. (Note
that a different definition of the dissociation temperature was used in
\cite{Sun:2007fc}.) In Fig.~\ref{Fig:phase_diagram} the phase diagram at zero
isospin chemical potential is drawn with the dissociation temperatures of the
pions, the diquark and antidiquark indicated. We can see that our definition of
the dissociation temperature nicely coincides with the position of the BCS--BEC
crossover, to be discussed in more detail in the following section. It is
suitable to remark here that when the effects of confinement are taken into
account, for example by coupling the NJL model to the Polyakov loop, the
transition temperature becomes somewhat higher than in
Fig.~\ref{Fig:phase_diagram} \cite{Brauner:2009gu}. However, the qualitative
structure of the phase diagram is not changed. Also, it should be stressed that
adding the Polyakov loop would only modify the results of this section. It would
have no effect at all on the phase diagram in the $(\mu_B,\mu_I)$ plane at zero
temperature, discussed in Sec.~\ref{Sec:phase_diagram}, and it plays an
absolutely negligible role in the thermodynamics of diquark BEC, to be
investigated in Sec.~\ref{Sec:BEC} \cite{Hands:2006ve}.


\section{Thermodynamics of diquark Bose--Einstein condensation}
\label{Sec:BEC}
The main virtue of two-color QCD, providing a strong motivation for its study,
is the possibility of lattice Monte Carlo simulations. It is therefore of
utmost importance to try to match model calculations to available lattice data.
To this end, one should note that in order to have a positive determinant of the
Dirac operator that in turn makes the importance sampling possible, one needs an
even number of quark flavors with the same chemical potentials. The phase
diagram in the $(\mu_B,\mu_I)$ plane investigated in
Sec.~\ref{Sec:phase_diagram} is therefore not accessible to current lattice
techniques. The lattice literature on two-color QCD at nonzero density
invariably addresses the case of zero isospin chemical potential, where the
individual quark flavors indeed share the common chemical potential, $\mu_B$.
The latest simulations \cite{Hands:2006ve,*Hands:2010gd} study the thermodynamic
properties of two-color QCD with two quark flavors at low temperature as a
function of the baryon chemical potential. (They also calculate the gluon
propagator which, however, cannot be determined in model approaches based
on the flavor symmetry.)

\begin{figure}
\begin{center}
\includegraphics[scale=1]{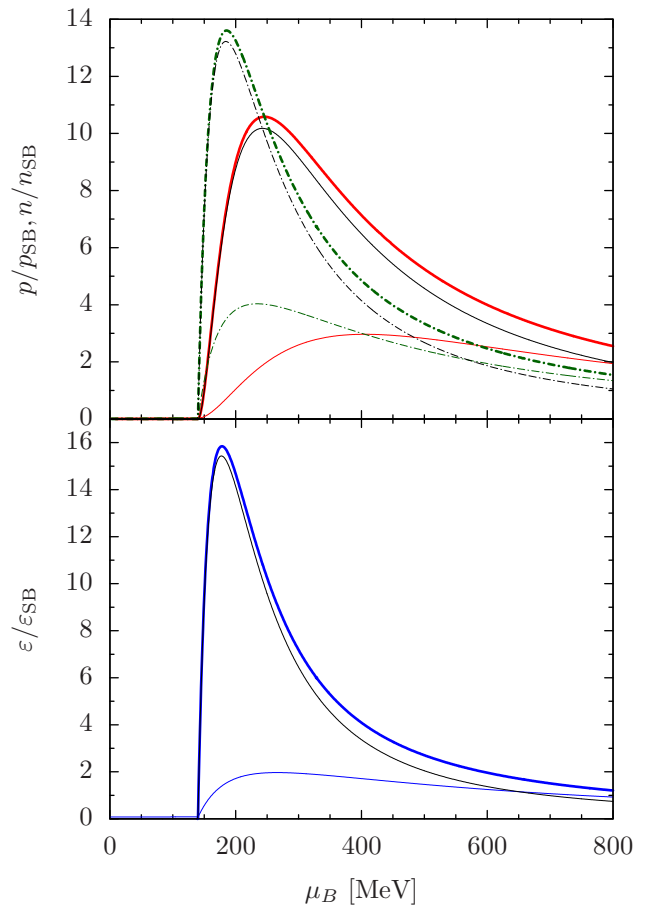}
\end{center}
\caption{Pressure and baryon density (upper panel) and energy density (lower
panel), normalized to the Stefan--Boltzmann values, at zero isospin chemical
potential as a function of baryon number chemical potential. The thick lines
correspond to $T=0\text{ MeV}$ with the following notation: the red solid line
for pressure, the green dash-dotted line for baryon density, and the blue solid
line for energy density. The thin lines with the same line notation show the
results for $T=50\text{ MeV}$. The thin black lines, almost coinciding with the
zero-temperature NJL values, are the results of chiral perturbation theory
at leading order.}
\label{Fig:thermodynamics}
\end{figure}
Within our NJL model, we calculate the pressure from the thermodynamic potential
(\ref{subtr_Omega}), making an appropriate constant subtraction to ensure that
the vacuum pressure is zero. The baryon number density and energy density are
in turn obtained from the pressure using the thermodynamic identities,
$n_B=\partial p/\partial\mu_B$ and $\varepsilon=-p+\mu_Bn_B+Ts$, where the
entropy density reads $s=\partial p/\partial T$. Since at high chemical
potentials the system is expected to behave as a weakly interacting Fermi gas,
it is convenient to normalize these thermodynamic quantities by the
Stefan--Boltzmann (SB) values, corresponding to a free massless Fermi gas.
Taking into account that two-color QCD with two quark flavors includes four
different species of Dirac fermions, these read
$$
p_{\text{SB}}=\frac{\mu_B^4}{48\pi^2}+\frac{\mu_B^2T^2}6+\frac{7\pi^2T^4}{45},
\quad
n_{\text{SB}}=\frac{\mu_B^3}{12\pi^2}+\frac{\mu_BT^2}3,
$$
and $\varepsilon_{\text{SB}}=3p_{\text{SB}}$. The numerical results for two
different temperatures are shown by the colored lines in
Fig.~\ref{Fig:thermodynamics} \footnote{In order to minimize the cutoff
artifacts in the NJL model calculation, we evalute the SB thermodynamic
quantities using the same model with the cutoff, setting the constituent quark
mass as well as other condensates to zero by hand. Using this prescription,
the SB-normalized expressions go to one---as they should---in the limit of
large chemical potential and/or temperature.}. Remarkably, even for the highest
chemical potentials displayed, the pressure and density are still rather far
from the SB limit. In the mean-field approximation, this is obviously due to
the pairing gap.

Since near the diquark BEC phase transition, the theory behaves like a dilute
Bose gas rather than a system of noninteracting quasiparticles, described by the
NJL model in the mean-field approximation, we calculate for comparison the same
quantities in a model with elementary scalar degrees of freedom. For the sake
of simplicity, we use the $\gr{O(6)}$ linear sigma model at tree level
(corresponding to zero temperature). This is defined by the Euclidean Lagrangian
\begin{equation}
\La_{\text{l$\sigma$m}}=\frac12(\partial_\mu\vec\phi)^2-\frac12M^2\vec\phi^2+
\frac14\lambda(\vec\phi^2)^2-H\sigma,
\label{lsm_lagr}
\end{equation}
$\vec\phi=(\sigma,\vec\pi,\mathrm{Re}\,\Delta,\mathrm{Im}\,\Delta)^T$, where
$\vec\pi$ is the pion triplet and $\Delta$ the complex diquark field. The
parameter $H$ breaks the $\gr{O(6)}$ symmetry explicitly down to $\gr{O(5)}$.
The baryon chemical potential $\mu_B$ is incorporated by the replacement of the
ordinary derivatives with covariant ones, acting nontrivially only on the
diquark, $D_0\Delta=(\partial_0-\mu_B)\Delta$, and
$D_0\he\Delta=(\partial_0+\mu_B)\he\Delta$.

The tree-level analysis of this model is straightforward, so we proceed
directly to the results. The negative mass squared triggers condensation
of the field, which, in the vacuum, is aligned with the explicit symmetry
breaking term, $\sigma\equiv f_\pi=\sqrt{(M^2+m_\pi^2)/\lambda}$, where the pion
mass is determined implicitly by $H=f_\pi m_\pi^2$. As soon as the baryon
chemical potential exceeds the pion mass, diquarks condense. The chiral
and diquark condensates are then in general parameterized by
$\sigma=v\cos\theta$ and $\Delta=v\sin\theta$. For the magnitude $v$ and the
angle $\theta$ one finds the expressions 
$$
v^2=\frac{M^2+\mu_B^2}\lambda,\quad
\cos\theta=\frac{m_\pi^2}{\mu_B^2}\sqrt{\frac{M^2+m_\pi^2}{M^2+\mu_B^2}}.
$$
The unphysical Lagrangian parameter $M^2$ may be eliminated in favor of the
mass of the ``radial'' excitation, conventionally identified with the sigma
particle, $m_\sigma^2=2M^2+3m_\pi^2$. All quantities of interest can then be
expressed in terms of observables. For instance, the diquark condensate is
given by
$$
\Delta=f_\pi\sqrt{1-\left(\frac{m_\pi}{\mu_B}\right)^4+
2\frac{\mu_B^2-m_\pi^2}{m_\sigma^2-m_\pi^2}}.
$$
Note that the results of chiral perturbation theory
\cite{Kogut:1999iv,*Kogut:2000ek} are recovered by taking the limit
$m_\sigma\to\infty$. Also, the expressions for the chiral condensate,
$\sigma=f_\pi m_\pi^2/\mu_B^2$, and the baryon density,
$n_B=\mu_B\Delta^2$, are actually identical to those in the
chiral perturbation theory. (The baryon density, nevertheless, takes on a
different value due to the modified value of the diquark condensate.)
It is interesting to observe that as the chemical potential approaches $m_\pi$
from the right, the predictions of the linear sigma model and chiral
perturbation theory are related by
$\Delta_{\text{l$\sigma$m}}/\Delta_{\chi\text{PT}}=m_\sigma/\sqrt{m_\sigma^2-
m_\pi^2}$. This means that as soon as the diquark condensate is nonzero, a
finite sigma mass drives the results off the predictions of chiral perturbation
theory even in the limit of zero density. Given the above solution of the gap
equations, it is straightforward to calculate the thermodynamic quantities.
Using again the normalization with respect to the SB values, one finds
\begin{align}
\notag
\frac{p_{\text{l$\sigma$m}}}{p_{\text{SB}}}&=
24\pi^2\left(\frac{f_\pi}{m_\pi}\right)^2
\left(1-\frac1{x^2}\right)^2
\left(\frac1{x^2}+\frac{1}{\tau^2-1}\right),\\
\label{nep_lsm}
\frac{n_{\text{l$\sigma$m}}}{n_{\text{SB}}}&=
\frac{12\pi^2}{x^2}\left(\frac{f_\pi}{m_\pi}\right)^2
\left(1-\frac1{x^4}+2\frac{x^2-1}{\tau^2-1}\right),\\
\notag
\frac{\varepsilon_{\text{l$\sigma$m}}}{\varepsilon_{\text{SB}}}&=
8\pi^2\left(\frac{f_\pi}{m_\pi}\right)^2
\frac{x^2-1}{x^4}
\left(1+\frac3{x^2}+\frac{3x^2+1}{\tau^2-1}\right),
\end{align}
where we used the shorthand notation $x=\mu_B/m_\pi$ and $\tau=m_\sigma/m_\pi$.
Similar expressions were obtained for pion condensation in
\cite{Campbell:1974qt,*He:2005nk}. It is amusing, though perhaps not very
physical, that all the expressions (\ref{nep_lsm}) have a nonzero limit as
$x\to\infty$. This takes even the same value for all three observables,
$24\pi^2f_\pi^2/(m_\sigma^2-m_\pi^2)$.

Taking the values of $m_\pi$ and $f_\pi$ over from the NJL model, these
predictions of the linear sigma model depend on a single parameter, namely
$\tau$. The parameter-free results in the limit $\tau\to\infty$ (that is,
chiral perturbation theory) are shown in Fig.~\ref{Fig:thermodynamics} by
the thin black lines. In fact, adjusting $\tau\approx9$, one would find that
the NJL and linear sigma model curves lie exactly on top of each other up to
$\mu_B$ about $400\text{ MeV}$, and remain very close in the whole range of
chemical potentials displayed in Fig.~\ref{Fig:thermodynamics}. A cautious
reader may wonder why this is so, remembering our claim in
Sec.~\ref{Sec:phase_diagram} that at $\mu_B\approx229\text{ MeV}$ a crossover
from the BEC behavior to quark matter occurs. The reason for this unexpected
agreement even after the crossover to quark matter, defined by the change of
sign of the expression $M-\mu_B/2$, is that the quark excitations will keep a
large energy gap. At low temperatures, they will not be excited and the bulk
thermodynamic quantities will be dominated by the condensate (and its
fluctuations, as we will see later).

Let us now compare our calculation based on chiral modes with the lattice data.
A glance at the results presented in
Ref.~\cite{Hands:2006ve,*Hands:2010gd} reveals that all three thermodynamic
observables considered here have substantially different values. This is, of
course, not so surprising on account of the fact that we made no attempt to fit
our model parameters to the lattice data. However, this issue deserves more
care, and we would now like to explain to what extent the lattice data
actually can be explained with chiral models.

First of all, one should note that the lattice computations are always
performed with a nonzero source for the diquark operator. While to obtain
physical predictions one should eventually take the limit of zero source, for
the sake of comparison it is technically much easier to include the
nonzero source in the model calculations. Indeed, in the NJL model, adding a
diquark source amounts simply to shifting the diquark field $\Delta$ in
the Yukawa interaction term in (\ref{Lagr_bosonized}), and consequently also in
the quasifermion contribution to the thermodynamic potential (\ref{TDpot}). In
the linear sigma model Lagrangian (\ref{lsm_lagr}), we would replace $H\sigma$
with $H\sigma+J\mathrm{Re}\,\Delta$, $J$ being the diquark source. The
resulting gap equations can still be solved analytically, albeit implicitly, in
the chiral perturbation theory limit, that is, $\tau\to\infty$
\cite{Kogut:1999iv,*Kogut:2000ek}. Writing the sources $H$ and $J$ in terms of a
new angle $\varphi$ as $H=f_\pi m_\pi^2\cos\varphi$ and $J=f_\pi
m_\pi^2\sin\varphi$, the gap equation for $\theta$, which determines the
orientation of the condensate, reads
$$
x^2\sin\theta\cos\theta=\sin(\theta-\varphi).
$$
Once this equation is solved for $\theta$, the thermodynamic observables are
obtained from the generalization of Eq.~(\ref{nep_lsm}) to nonzero source in the
limit $\tau\to\infty$,
\begin{align*}
\frac{p_{\chi\text{PT}}}{p_{\text{SB}}}&=
48\pi^2\left(\frac{f_\pi}{m_\pi}\right)^2
\frac1{x^4}\left[\frac12x^2\sin^2\theta+
\cos(\theta-\varphi)-1\right],\\
\frac{n_{\chi\text{PT}}}{n_{\text{SB}}}&=
\frac{12\pi^2}{x^2}\left(\frac{f_\pi}{m_\pi}\right)^2
\sin^2\theta,\\
\frac{\varepsilon_{\chi\text{PT}}}{\varepsilon_{\text{SB}}}&=
16\pi^2\left(\frac{f_\pi}{m_\pi}\right)^2
\frac{1}{x^4}
\left[\frac12x^2\sin^2\theta-
\cos(\theta-\varphi)+1\right].
\end{align*}

\begin{figure}
\begin{center}
\includegraphics[scale=1]{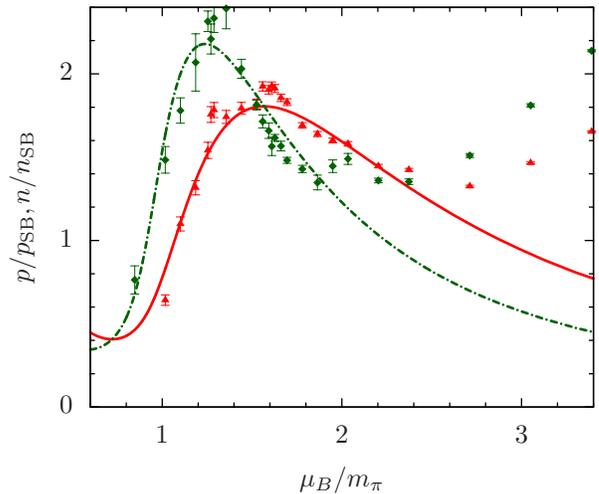}
\end{center}
\caption{Comparison of lattice results \cite{Hands:2006ve} (we are indebted to
S.~Hands for providing us with the data) and leading order chiral perturbation
theory with nonzero diquark source. The line notation is the same as in
Fig.~\ref{Fig:thermodynamics}: red solid for pressure and green dash-dotted for
baryon density.}
\label{Fig:CHPTsource}
\end{figure}
The results of a sample calculation using chiral perturbation theory with a
nonzero diquark source are shown in Fig.~\ref{Fig:CHPTsource}. Since we were
not able to fix the input parameters in the vacuum, we simply fit the prediction
of chiral perturbation theory to lattice results at nonzero chemical potential.
We thus obtain $\varphi\approx0.10$, $f_\pi/m_\pi\approx0.21$, and
$m_\pi\approx0.59$ in lattice units. (We used only the data for baryon density
in the fit, since the pressure may get a nontrivial contribution from the gluon
sector of QCD. Also, only the points with $\mu_B/m_\pi\lesssim2$ were taken
into account.) Doing this, we merely wish to demonstrate that a part of the
lattice data can be explained using a model based just on the chiral symmetry.
Obviously, the same numerical values can also be obtained with the NJL model; we
did not do that since it would require tuning an even larger number of
parameters at will.

It is clear that the high numerical values of the NJL prediction plotted in
Fig.~\ref{Fig:thermodynamics} do not present a problem. In fact,
Eq.~(\ref{nep_lsm}) illustrates convincingly that they scale sensitively with
the ratio $f_\pi/m_\pi$. What is a problem, however, is the large ratio of the
heights of the peaks in the energy density, and in the pressure and baryon
density, as observed on the lattice \cite{Hands:2006ve,*Hands:2010gd}. This is
the reason why we did not plot the energy density in Fig.~\ref{Fig:CHPTsource}
at all. Both the mean-field NJL model and the linear sigma model at tree level
(and in turn the chiral perturbation theory at the leading order) predict all
three peaks to have heights of roughly the same magnitude. Even though we
cannot offer and ultimate resolution of this discrepancy, we propose a possible
explanation.

Our reasoning goes back to the thermodynamic identity $\varepsilon=-p+\mu_B
n_B+Ts$. Note that the lattice simulations were performed at low
temperature, roughly an order of magnitude smaller than the typical
values of the chemical potential involved. The SB expressions
are then well approximated by their zero-temperature limits, which satisfy the
relation
$\varepsilon_{\text{SB}}^0=3p_{\text{SB}}^0=\frac34\mu_Bn_{\text{SB}}^0$. The
SB-normalized thermodynamic quantities (denoted by a bar) thus satisfy the
identity
$$
\bar\varepsilon\approx-\frac13\bar p+\frac43\bar
n_B+\frac{Ts}{\varepsilon_{\text{SB}}},
$$
where the approximation sign refers to the zero-temperature limit of the SB
values. Using this identity, one may from the values of $\bar
p,\bar n_B,\bar\varepsilon$ given in \cite{Hands:2006ve,*Hands:2010gd}
determine the purely thermal component of the energy density, represented by
the entropy. One thus finds that $Ts/\varepsilon_{\text{SB}}$ is essentially
zero for $\mu_B/m_\pi\gtrsim3$ \footnote{This suggests that the Karsch
coefficient, that renormalizes the energy density without affecting the
pressure and baryon density and is independent of the chemical potential
\cite{Hands:2005yq,*Hands:2006ve,*Hands:2010gd}, should not be much smaller than
one. Otherwise, the entropy would come out negative at high chemical potential,
which would, of course, be unphysical.}, but the sharp peak already present in
$\bar\varepsilon$, of course, remains. We are thus led to the conclusion that
the high peak in the energy density is due to thermal excitations.

In order to see how this can come about, let us observe that the diquark
dispersion relation below the onset of BEC is given by $E_{\vek k}=\sqrt{\vek
k^2+m_\pi^2}-\mu_B$. Right at the BEC transition point, $\mu_B=m_\pi$, the
dispersion is therefore quadratic at low momentum, and it still remains rather
flat even when the chemical potential increases above $m_\pi$. Moreover, in the
whole diquark condensation phase there is a massless NG boson stemming from the
spontaneous breaking of the exact global symmetry associated with the
conservation of the baryon number. As soon as the temperature is nonzero, the
NG states will be abundantly populated, giving a large entropy, contributing in
turn to the energy density. Whether this mechanism is physically correct can
easily be tested by changing the temperature. If the peak in the energy density
is really mostly due to thermal excitations, its height should be very
sensitive to the changes in temperature. On the other hand, the peaks in the
pressure and baryon density, not getting a contribution from the entropy,
should be rather robust, being dominated by the condensate contribution.

With this physical picture in mind, it is now also clear how our model
calculations should be modified in order to take the NG excitations in the
account. In the linear sigma model or chiral perturbation theory, one must
include loop effects which introduce the finite temperature into the system via
the Matsubara summation. Within the NJL model, the mean-field approximation
only includes quasifermionic modes. In order that the NG excitations be
properly taken into account, one must, again, go beyond the mean-field
approximation. Although both these extensions are in principle straightforward
(the next-to-leading order results within chiral perturbation theory being even
already available \cite{Splittorff:2001fy,*Splittorff:2002xn}), albeit maybe
somewhat tedious, they may still not yield the desired quantitative agreement
with the lattice simulations. Possible reasons are discussed in the concluding
section.


\section{Conclusions}
\label{Sec:conclusions}
In this paper we investigated the thermodynamics of two-color QCD with two
light quark flavors. We devised a general method to construct interaction
Lagrangians of the NJL type with the extended global $\gr{SU(4)}$ symmetry.
Using the mean-field approximation, we then determined the phase diagram in the
$(\mu_B,\mu_I)$ plane, including the FF pairing and the isospin splitting of
the flavor chiral condensates.

One should emphasize that, since we only considered the chiral NJL model, 
confinement effects are not taken into account in our results. This can be done
phenomenologically by adding the Polyakov loop as an additional degree of
freedom \cite{Brauner:2009gu}. However, this will only affect the physics at
nonzero temperature, since the Polyakov loop NJL model reduces to pure NJL at
zero temperature. Hence, the results of Sec.~\ref{Sec:phase_diagram} will be
unaffected. Likewise, Figures \ref{Fig:masses}, \ref{Fig:spectral}, and
\ref{Fig:phase_diagram} will not change qualitatively; basically the only
difference will be a rescaling of the critical temperatures to higher values.

Finally, in Sec.~\ref{Sec:BEC} we discussed the thermodynamics of diquark BEC
with a particular attention to available lattice data. We suggested that a
proper inclusion of thermal NG excitations is crucial in order to achieve a
quantitative agreement with them. While it would be very nice to pursue our
model calculations as far as possible, we have some reservations as to whether
this would lead to the desired goal. The reason is that the lattice
computations were performed with rather heavy quarks. Consequently, the
lightest non-NG bosonic state is the rho meson with $m_\pi/m_\rho\approx0.8$.
We see that there is no clear separation of scales, and therefore the success of
any approach based on chiral symmetry and its spontaneous breaking is a
priori questionable. One obvious remedy to this situation would be
to perform lattice simulations with lighter quarks, but this may be technically
rather demanding. On the other hand, one might try to avoid using chiral models
by resorting to first-principle QCD calculations, based, e.g., on the
Dyson--Schwinger equations \cite{Maris:1997eg,*Nickel:2006vf,*Klahn:2009mb} or
the functional renormalization group approach 
\cite{Berges:1998sd,*Schaefer:2006sr}.

Summarizing, we would like to stress once more that two-color QCD is a system
whose study may teach about deconfinement in cold dense matter as well as
other aspects of the real world. At present, the techniques that would allow us
to investigate the phase diagram of two-color QCD in detail are on the market.
We believe that our future efforts will contribute to its general understanding.


\begin{acknowledgments}
The authors are grateful to H.~Abuki, K.~Fukushima, S.~Hands, X.-g.~Huang,
A.~Maas, D.~Nicmorus, D.~Parganlija, and H.~Warringa for useful discussions and
correspondence. The hospitality of the Niels Bohr International Academy where
this project was initiated is gratefully appreciated. Part of the work was
carried out during the stay of T.B. at the Norwegian University of Science and
Technology, Trondheim. The research of T.B. was supported in part by the
Alexander von Humboldt Foundation, and by the ExtreMe Matter Institute EMMI in
the framework of the Helmholtz Alliance Program of the Helmholtz Association
(HA216/EMMI). Numerical calculations were performed using the facilities of
the Frankfurt Center for Scientific Computing.
\end{acknowledgments}


\appendix
\section*{Appendix: Thermodynamic potential for Fulde--Ferrell pairing}
\label{App:LOFF}
Consider for simplicity a nonrelativistic theory of two fermion species,
$\psi_\uparrow,\psi_\downarrow$, defined by the (Euclidean) Lagrangian
$$
\La=\sum_{\sigma=\uparrow,\downarrow}
\he\psi_\sigma\left(\de_\tau-\frac{\vek\nabla^2}{2m}
-\hat\mu\right)\psi_\sigma
-g\he\psi_\uparrow\he\psi_\downarrow\psi_\downarrow\psi_\uparrow,
$$
where $g$ is the bare (attractive) coupling and $\hat\mu$ the diagonal chemical
potential matrix in the flavor space. Introducing a collective field
$\phi\sim\psi_\downarrow\psi_\uparrow$, performing the Hubbard--Stratonovich
transformation, integrating out the fermions, and finally summing over the
Matsubara frequencies, one arrives at an expression for the mean-field
thermodynamic potential analogous to Eq.~(\ref{TDpot}),
\begin{equation}
\begin{split}
\Omega[\phi]=&\int d^3\vek x\,\frac{|\phi(\vek x)|^2}g\\
&-\tr\left\{\frac12\Ha(\vek x)+T\log\left[1+e^{-\beta\Ha(\vek
x)}\right]\right\},
\end{split}
\label{NR_TDpot}
\end{equation}
where
\begin{equation*}
\Ha(\vek x)=\begin{pmatrix}
-\frac{\vek\nabla^2}{2m}-\mu_\uparrow & -\phi(\vek x)\\
-\phi^*(\vek x) & \frac{\vek\nabla^2}{2m}+\mu_\downarrow
\end{pmatrix},
\end{equation*}
is the matrix Hamiltonian in the Nambu space,
$(\psi_\uparrow,\he\psi_\downarrow)$, and the trace operation is to be
performed in the operator sense.

So far, no assumption on the particular form of the mean field configuration
$\phi(\vek x)$ was made; in principle this is to be determined by minimization
of the \emph{functional} $\Omega[\phi]$. Once again, however,
Eq.~(\ref{NR_TDpot}) is only formal because $\tr\Ha(\vek x)$ is badly
divergent. Subtracting the same expression for the normal phase (that is,
$\phi=0$) at zero temperature with the corresponding Hamiltonian $\Ha_0$, one
obtains the regularized thermodynamic potential as
\begin{equation}
\begin{split}
\Omega_{\text{sub}}[\phi]=&\int d^3\vek x\,\frac{|\phi(\vek x)|^2}g-
\tr\biggl\{\frac12\bigl[\Ha(\vek x)-|\Ha_0|\bigr]\\
&+T\log\left[1+e^{-\beta\Ha(\vek x)}\right]\biggr\}.
\end{split}
\label{NR_TDpot_sub}
\end{equation}
Within the nonrelativistic theory this prescription can be shown to remove all
divergences once the coupling $g$ is renormalized. (In practice this is usually
done by trading $g$ for the gap at zero temperature which is physically
observable and thus free of divergences.)

Unfortunately, it is impossible to evaluate the trace of the log term for a
general field configuration $\phi(\vek x)$. The FF pairing corresponds to the
Ansatz $\phi(\vek x)=\Delta e^{2i\vek q\cdot\vek x}$. Now the Hamiltonian may
already be diagonalized. To this end, we perform the unitary transformation,
$\Ha\to\Ha'=\he{U_{\vek q}}\Ha U_{\vek q}$, where $U_{\vek
q}=\mathrm{diag}\,(e^{i\vek q\cdot\vek x},e^{-i\vek q\cdot\vek x})$ and
$$
\Ha'=
\begin{pmatrix}
\frac{(-i\vek\nabla+\vek q)^2}{2m}-\mu_\uparrow & -\Delta\\
-\Delta & -\frac{(-i\vek\nabla-\vek q)^2}{2m}+\mu_\downarrow
\end{pmatrix}.
$$
The eigenvalues of this Hamiltonian are now easily found by Fourier
transforming to momentum space. They are $E^-_{\vek k}$ and $-E^+_{\vek k}$,
where $E^\pm_{\vek k}=E_{\vek k}\pm\delta\mu_{\vek q}$, $\delta\mu_{\vek
q}=\delta\mu-(\vek k\cdot\vek q)/m$, $E_{\vek k}=\sqrt{\xi_{\vek
k}^2+\Delta^2}$, $\xi_{\vek k}=(\vek k^2+\vek q^2)/2m-\bar\mu$, and we have
defined $\bar\mu=(\mu_\uparrow+\mu_\downarrow)/2$ and
$\delta\mu=(\mu_\uparrow-\mu_\downarrow)/2$. Our crucial observation is that
\emph{since the trace in \eqref{NR_TDpot_sub} has a well defined mathematical
meaning in combination with $\Ha_0$, we have to do the same unitary
transformation $U_{\vek q}$ on it as well}. The thermodynamic potential density
therefore becomes
\begin{equation*}
\begin{split}
\frac{\Omega(\Delta,q)}V=&\frac{\Delta^2}g-\int\frac{d^3\vek k}{(2\pi)^3}
\biggl\{E_{\vek k}-\frac12\bigl[|\xi_{\vek k}+\delta\mu_{\vek q}|\\
&+|\xi_{\vek k}-\delta\mu_{\vek q}|\bigr]
+T\sum_{\pm}\log\left[1+e^{-\beta(E_{\vek k}\pm
\delta\mu_{\vek q})}\right]\biggr\}.
\end{split}
\end{equation*}
Note that subtracting naively the contribution of the normal phase from
Eq.~(\ref{NR_TDpot}) would amount to replacing the second term in the curly
brackets with $\frac12\bigl[|\xi_{\vek k}+\delta\mu|+|\xi_{\vek
k}-\delta\mu|\bigr]$. The difference is finite, but the naive subtraction would
lead to an extra term in the thermodynamic potential, independent of $\Delta$
and going to $-\infty$ at large $\vek q$. Such a thermodynamic potential would
not even have a minimum. Our derivation makes sure that the thermodynamic
potential is both well defined and physically consistent.

In relativistic systems the same argument can be used, but the situation is
less simple. The reason is that the divergences in the relativistic theory are
more severe due to different quasiparticle dispersion relation, which is
already signalized by the presence of a quadratic divergence
$\sim\Lambda^2q^2$. The divergences are not completely removed by the
subtraction of the normal phase and renormalization of the coupling constant.
This means that the subtraction procedure remains somewhat ambiguous and is to
be understood as a part of the definition of the model. This is, however,
nothing new in the NJL model where one often handles divergent quantities in a
rather formal manner \cite{Klevansky:1992qe}. Our strategy is as follows. We
define the naive thermodynamic potential, $\Omega(\sigma,\rho,\Delta(\vek x))$,
analogously to Eq.~(\ref{NR_TDpot}), and rewrite it as
$$
\Omega(\sigma,\rho,\Delta(\vek x))=\Omega(\sigma,\rho,\Delta(\vek x))-
\Omega(\sigma,\rho,0)+\Omega(\sigma,\rho,0).
$$
The last term is the thermodynamic potential of the ``normal phase''
while in the difference of the first two terms the most severe divergences
cancel. It is this difference that the unitary transformation $U_{\vek q}$ is
applied to. For the FF pairing this procedure leads to Eq.~(\ref{subtr_Omega}).
One should, however, observe that since the final expression is still mildly
(logarithmically) divergent, the result depends on the choice of the term to
subtract, which was already noted in \cite{Fukushima:2007bj}.

Here we have subtracted the ``normal phase'' with respect to the diquark
condensate, but with unchanged values of the chiral condensates. By dimensional
analysis, the thermodynamic potential suffers from the following $\vek
q$-dependent divergences: a quadratic divergence $\sim q^2$ and logarithmic
divergences $\sim q^2\sigma^2,q^2\rho^2,q^2\mu_{B,I}^2,q^2\Delta^2$. All
divergences but the very last one are canceled by our subtraction procedure.
Fortunately, the remaining divergence $\sim q^2\Delta^2$ is rather mild because
in the FF phase the gap $\Delta$ is numerically much smaller than in the BCS
phase.


\bibliography{personal}

\end{document}